\documentclass[%
 reprint,
 amsmath,amssymb,
 aps,
]{revtex4-2}

\usepackage{graphicx}
\usepackage{dcolumn}
\usepackage{bm}
\usepackage{color, soul}

\usepackage{color}
\usepackage{ulem}
\def\a{s}
\newcommand{\showc}[1]{\if\a\b{{\hl{#1}}}\else{#1}\fi}
\newcommand{\add}[1]{\if\a\b{{\color{red} #1}}\else{#1}\fi}
\newcommand{\comm}[1]{\if\a\b{{\color{blue}\{\small \sc #1\}}}\else{}\fi}
\newcommand{\del}[1]{{\if\a\b{{\color{magenta}\sout {#1}}}\else{}\fi}}

\def\sfc{\chi}

\begin{document}
\preprint{APS/123-QED}

\title{Optomechanical Self-Stability of Freestanding Photonic Metasurfaces}

\author{Avinash Kumar}
\author{Daniel Kindem}
\author{Ognjen Ilic*}%

\affiliation{%
 Department of Mechanical Engineering, University of Minnesota, Minneapolis, MN 55455, USA
}%


\begin{abstract}
We develop an analytical framework for self-stabilizing optical manipulation of freestanding metasurfaces in three dimensions. Our framework reveals that the challenging problem of stabilization against translational and rotational perturbations in three dimensions is reduced to a simpler scattering analysis of the metasurface unit cell in two dimensions. We derive universal analytical stiffness coefficients applicable to arbitrary three-dimensional radial metasurfaces and radial beam intensity profiles. The analytical nature of our framework facilitates highly efficient discovery of optimal optomechanical metasurfaces. Such use of metasurfaces for mechanical stabilization enables macroscale and long-range control in collimated, but otherwise unfocused light beams, and could open up new avenues for manipulation beyond traditional optical tweezing and transport.
\end{abstract}

\maketitle

\section{Introduction}
The use of light for contactless mechanical manipulation of freestanding objects spans a wide and diverse set of applications across biology and biomedicine \cite{Fazal_2011, Stevenson_2010, Dholakia_2011}, colloidal science \cite{Grier_1997, Kumar_2013, Martinez_2017} and microfluidics \cite{MacDonald_2003, Padgett_2011a, Mohanty_2012}, as well as chemistry \cite{Kitamura_2003, Moffitt_2008, Zemanek_2019}, and quantum optomechanics \cite{Chang_2009, Romero_Isart_2010, Li_2011, Gieseler_2012, Neukirch_2015, Bhattacharya_2017}. In typical approaches to optical manipulation---such as optical tweezing and transport \cite{Ashkin_1970, Grier_2003, Baumgartl_2008, Padgett_2011, Brzobohaty_2013, Taylor_2015}---particles and small objects are trapped by strong optical field gradients created by focusing light into a target spot. This need for focused light can limit the size of objects that can be manipulated, as well as the volume of space and the distance at which manipulation is effective. In contrast, manipulation with collimated, unfocused beams could overcome these limitations but is inherently unstable: a slight disturbance of an object away from the beam axis results in destabilizing radiation pressure. Efforts to tackle the problem of stabilizing freestanding objects have relied on prescribing a particular geometric shape, such as a parabolic \cite{Popova_2016} or spherical \cite{Manchester_2017}, with the intention of inducing counter-balancing optical forces and torques. Such approaches introduce further challenges of fabricating structures with precise three-dimensional shape. In contrast, nanostructured interfaces offer a means of controlling the optical force \cite{Srinivasan_2016a, Swartzlander_2017, Ilic_2018a, Chu_2018, Achouri_2019, Chu_2019}, leading to concepts of passively restoring optical manipulation ~\cite{Ilic_2019, Siegel_2019, Srivastava_2019, Srivastava_2020c, Salary_2020}. However, examples in the literature have been limited to sub-optimal photonic designs, restrictions to two-dimensional models, and the need for additional parasitic tethered masses to offset the center of mass for stability. Crucially, there has been no approach to analytically assess self-stability of nanostructured objects and their dynamics in three dimensions.

In this work, we develop a general optomechanical stabilization framework that is analytical and applicable for dynamics in three dimensions, and that further facilitates discovery of non-conventional but optimal metasurfaces. We use symmetry and perturbation calculus to demonstrate that the challenging problem of stability in three dimensions can be formulated as a much simpler scattering analysis of the metasurface unit cell in two dimensions in equilibrium. We establish universal expressions for analytical stiffness coefficients, applicable to arbitrary radial metasurface elements and radial beam intensity profiles. Due to its analytical nature, our framework facilitates efficient global exploration of optimal metasurfaces and also beam intensity profiles, subject to conditions for stabilization in three dimensions. As we show by example, a number of relevant design figures of merit---e.g., maximizing force/torque, stiffness, beam power utilization, etc.---can be incorporated in a straightforward manner for effective optimization and refinement over a broad parameter space.

\section{Methods and Results}
\subsection{Analytical formalism for metasurface self-stability}
\begin{figure}[t]
\centering
    \includegraphics[width=85mm]{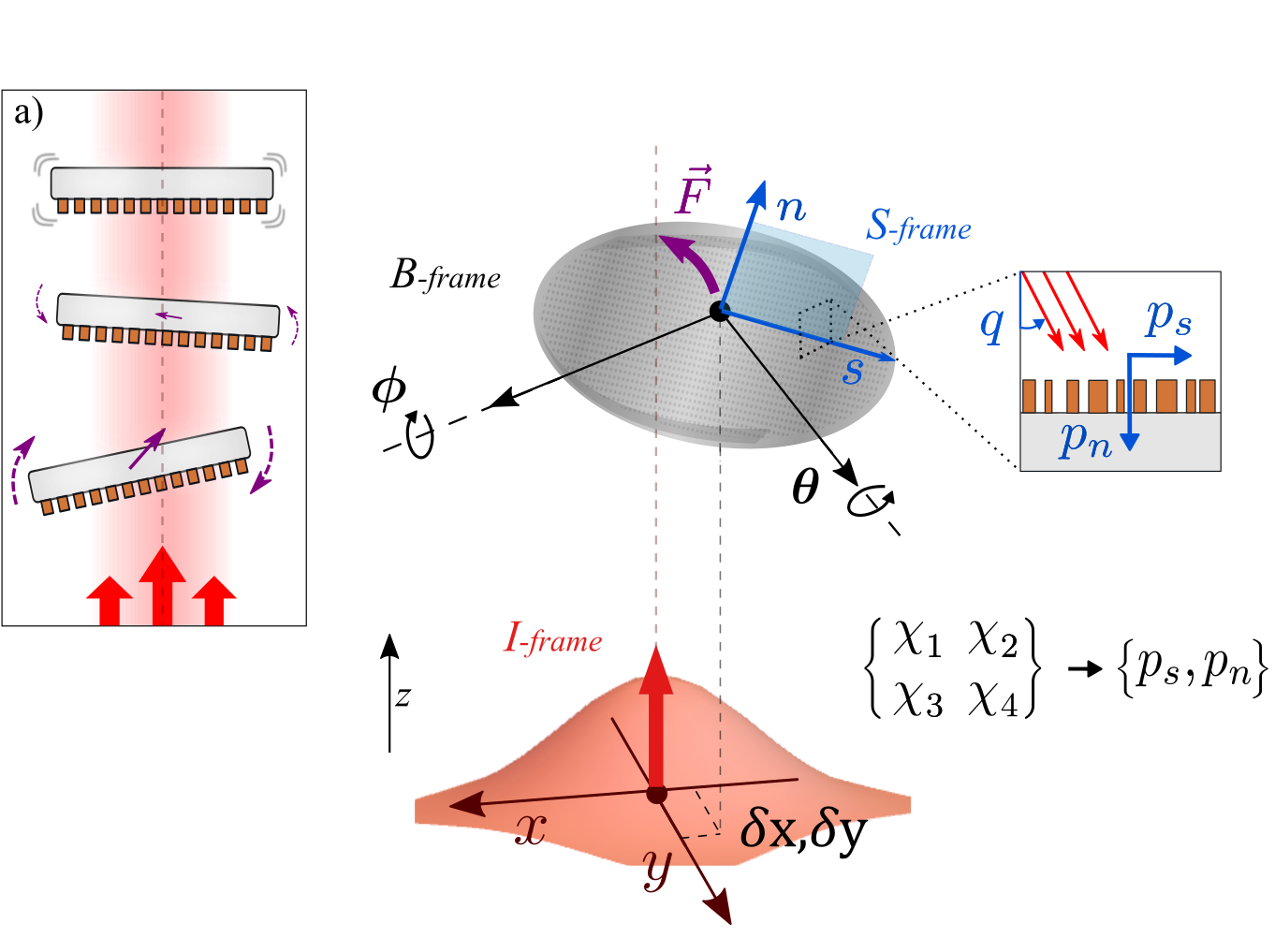}
    \caption{\label{fig:schematic_unitcell} Optomechanical stabilization of freestanding objects with a nanostructured metasurface (inset: cross-sectional view). Embedded, radially varying, building block elements can generate passively restoring optical forces $\vec{F}$ when the metasurface is displaced from its equilibrium position on the beam axis. For lateral translations and rotations relative to the axis of an arbitrary varying beam, the three-dimensional opto-mechanical response of the object can be fully captured by the normal ($p_{n}$) and the radial ($p_{s}$) radiation pressure components in the two-dimensional plane of the unit cell (Eq.~(\ref{eq:fs})).}
\end{figure}

We consider a configuration for optomechanical manipulation where a beam impinges on a planar, freestanding three-dimensional object with a structured surface and size much larger than wavelength of light (Fig. \ref{fig:schematic_unitcell}). The surface of the object contains embedded, radially varying building-block elements that induce longitudinal and radial radiation pressure (Fig. \ref{fig:schematic_unitcell}, right inset). Our approach is general in that we make no assumptions about the photonic nature of these elements: for example, these could be unit cells of phase gradient metasurfaces \cite{Yu_2011, Aieta_2012, Kildishev_2013, Monticone_2013, Lin_2014a, Yu_2014, Arbabi_2017, Genevet_2017, Kamali_2018}, periodic Bloch-wave meta-gratings and photonic crystals \cite{Joannopoulos_2008, Fattal_2010}, metasurfaces based on anisotropic Mie scatterers \cite{Cihan_2018, Kuznetsov_2016}, or even a combination of two or more photonic motifs. In the frame of the metasurface, the impinging laser beam induces the spatially dependent radiation pressure components $p_{s,n}$ at each point on the surface. Depending on the position and the orientation of the object, it will experience a cumulative sum of forces and torques induced by its subelements. Specifically, the net force on the object, transformed into the inertial frame of the laser beam (frame $I$), is given by
\begin{gather}
\begin{bmatrix}
F_{x}  \\
F_{y}  \\
F_{z}  \\
\end{bmatrix}^{I}
= 
\int^{\frac{D}{2}}_{0} ds \int^{2 \pi}_{0} d\beta \ H_{S}^{I} \left (\beta,\theta,\phi \right)\begin{bmatrix}
p_{s} \\
0 \\
p_{n}  \\
\end{bmatrix} I( r_{\perp}) \cos(G) s,
\label{eq:F}
\end{gather}
where $s, \beta$ are the radial coordinate and the axial angle of the unit element, respectively, and $D$ is the overall length of the metasurface object. In this analysis, we assume that the size of the object and the spatial variation of the beam intensity are both much larger than the metasurface unit cell. For convenience of notation, the pressure components $p_{s,n}$ are normalized to the (local) light intensity $I$ and the speed of light $c$ (Appendix A). 

The transformation of the force from the metasurface unit cell frame $S$ to the laser frame $I$ is facilitated by the direction cosine matrix $H_S^I(\beta, \theta, \phi)$, where $\beta, \theta, \phi$ are the Euler angles describing the rotation of the object in three dimensions (Appendix A). In Eq.~(\ref{eq:F}), the beam intensity $I$ is evaluated at the radial distance $r_{\perp}$ away from the beam axis in the $I$-frame. This is calculated from the absolute position of the unit element in frame $I$---we label this position as $\bar{r}$. When the center of mass is laterally offset by distance $x$ and $y$ from the beam axis (in $I$-frame), we deduce:
\begin{gather}
\bar{r} =  
\begin{bmatrix}
x\\
y \\
0  \\
\end{bmatrix} 
+ H_S^I(\beta,\theta,\phi) 
\begin{bmatrix}
s\\
0 \\
0  \\
\end{bmatrix}
\label{eq:rbar}
\end{gather}
The radial distance in Eq.~(\ref{eq:F}) is then given by $r_{\perp} = \sqrt{ \bar{r}_x^2 + \bar{r}_y^2}$. Analogous to the force expressions, the torque experienced by the metasurface is given by
\begin{eqnarray}
\begin{bmatrix}
\tau_{x}  \\
\tau_{y}  \\
\tau_{z}  \\
\end{bmatrix}
= 
\int^{\frac{D}{2}}_{0} ds&&  \int^{2 \pi}_{0} d\beta \ 
\left ( H_{S}^{B} \begin{bmatrix}
s\\
0 \\
0  \\
\end{bmatrix} \right ) \nonumber \\
&&\times 
\left ( H_{S}^{B} \begin{bmatrix}
p_{s} \\
0 \\
p_{n}   \\
\end{bmatrix} \right )
I ( r_{\perp}) \cos(G) s
\label{eq:tau}
\end{eqnarray}
where $H_{S}^{B}$ is the coordinate transformation from the unit cell coordinate frame to the body axis coordinate frame (i.e., it is a rotation by angle $\beta$, namely $H_{S}^{B} = H_{S}^{I}(\beta,0,0)$). In both Eqs. ({\ref{eq:F}, \ref{eq:tau}), the $\cos(G)$ factor accounts for the projected area of the rotated object, i.e. the cosine of the angle between the incident wavevector and the object surface normal, namely $\cos(G) = \vec{k}_i^S \cdot \hat{n}$. The last transformation that is needed is the expression for $\vec{k}_i^S$, the incident beam wavevector in frame $S$. This is given by $\vec{k}_i^S = [k_s, k_t, k_n]^S = H_I^S \vec{k}_0$, where in the beam frame ($I$-frame) we assume incident light in the $z$ direction, $\vec{k}_0 || \hat{z}$ (see Appendix A for details). With these expressions in place, we can write the projected incident angle $q$, from Fig. \ref{fig:schematic_unitcell}, as $q = \tan^{-1} (k_s / k_n)$. The projected angle $q$ plays a key role in the stability analysis, as we show below.

We seek a general analytical framework to describe stabilization for arbitrary metasurfaces and beam profiles. For the metasurface to be stabilizing, it should seek to restore its position when displaced from its equilibrium on the beam axis. In general, the behavior of the metasurface is described by the nonlinear rigid-body equations of motion (Appendix A, Eq.~\ref{eq:EOM}). Here, we employ a perturbative analysis to describe the metasurface dynamics in the vicinity of the beam axis. Under the assumption of small displacements and small Euler angles, the translational motion and the rotational motion are coupled to first order, e.g., $\Ddot{x} = \frac{1}{m}\frac{\partial F_{x}}{\partial x}x + \frac{1}{m}\frac{\partial F_{x}}{\partial \theta}\theta$, and $\Ddot{\theta} = \frac{1}{\mathrm{I}_y}\frac{\partial \tau_{y}}{\partial x}x + \frac{1}{\mathrm{I}_y}\frac{\partial \tau_{y}}{\partial \theta}\theta$ (and similar for other coordinates), where $m$ is the metasurface mass and $\mathrm{I}_y$ is the moment of inertia about the $y$-axis. The general description of the dynamical system is given in Eq. (\ref{eq:dynsys}). For the purposes of assessing stability, the key information is provided by the Jacobian matrix of the system $\mathbf{J}$, whose elements are given by
\begin{equation}
    \mathbf{J}_{ij} = \frac{\partial f_i}{\partial u_j}
\end{equation}
where $\textit{\textbf{f}}$ relates to the forces/torques present in the system ($F_x, \tau_x$, etc.) and $\textit{\textbf{u}}$ the system variables ($x, \theta$, etc.). For the system analyzed in this work, the non-zero elements of the Jacobian matrix are (Appendix A)
\begin{eqnarray}
  \label{eq:sfc}
   \sfc_{1} &&= -\frac{1}{m} \frac{\partial F_{x}}{\partial x} = -\frac{1}{m} \frac{\partial F_{y}}{\partial y} \nonumber \\
   \sfc_{2} &&= \frac{1}{m} \frac{\partial F_{x}}{\partial \theta} = -\frac{1}{m} \frac{\partial F_{y}}{\partial \phi}  \nonumber \\
  \sfc_{3} &&= \frac{1}{\mathrm{I}_{y}} \frac{\partial \tau_{y}}{\partial x} = -\frac{1}{\mathrm{I}_{x}} \frac{\partial \tau_{x}}{\partial y}  \\
  \sfc_{4} &&= \frac{1}{\mathrm{I}_{x}} \frac{\partial \tau_{x}}{\partial \phi} = \frac{1}{\mathrm{I}_y} \frac{\partial \tau_{y}}{\partial \theta} \nonumber
\end{eqnarray}
where we explicitly define the \textit{stiffness} coefficients $\sfc_{1-4}$. For metasurface stabilization, there should be no eigenvalues of the Jacobian matrix with a positive real part. This condition can be stated as a set of \textit{stability conditions}  $c_i (\sfc_1,\sfc_2,\sfc_3,\sfc_4 )<0$ that are presented in Eq.~(\ref{eq:Cs}). We remark that despite preferential coupling of $x$-$\theta$ and $y$-$\phi$ in Eqs.~({\ref{eq:sfc}}), expressions for $\sfc_{1-4}$ must derive from the full three-dimensional dynamical behavior of the object: e.g., even when the motion along a single axis is considered, both the metasurface and the light scattered from it need to be treated as three-dimensional entities.

The first key result of this work is to demonstrate that the optomechanical response of a complex dynamical system of a translating/rotating metasurface in a light field can be embodied by a simpler subspace of unit cell radiation pressures. Specifically, we develop analytical expressions for the stiffness coefficients $\sfc_{1-4}$ of the system dependent on the radial $p_s$ and longitudinal $p_n$ radiation pressure components of the unit cell, where
\begin{subequations}
\label{eq:fs}
\begin{alignat}{4}
  \sfc_{1} &=  -\frac{\pi}{mc} \int_{0}^{D/2} p_{s} I'(s) s \ \mathrm{d}s \ \label{eq:f1} \\ 
  \sfc_{2} &= \frac{\pi}{mc} \int_{0}^{D/2} \left[- \frac{\partial p_{s}}{\partial q} + 2 p_{n}\right ] I(s) s \ \mathrm{d}s  \label{eq:f2} \\ 
  \sfc_{3} &=   -\frac{\pi}{\mathrm{I}c} \int_{0}^{D/2} p_{n} I'(s) s^2 \ \mathrm{d}s  \label{eq:f3} \\ 
  \sfc_{4} &= \frac{\pi}{\mathrm{I}c} \int_{0}^{D/2} \frac{\partial p_{n}}{\partial q}  I(s) s^2 \ \mathrm{d}s  \label{eq:f4} 
\end{alignat}
\end{subequations}
evaluated at equilibrium on the beam axis (see Appendix A for the complete derivation). Here, $s,n$ are the radial and the normal coordinate of the unit cell, respectively, and $m, \mathrm{I}$ are the mass and the (diagonal) moment of inertia of the metasurface of size $D$ (with $\mathrm{I}=\mathrm{I}_{x}=\mathrm{I}_{y}$), and $c$ is the speed of light. The profile of the laser beam is accounted for by expressions for the radial beam intensity $I(s)$ and the radial derivative of intensity $I'(s)= \partial I(s) / \partial s$. The response of the unit cell gives rise to the radial $p_s$ and the normal $p_n$ radiation pressures and derivatives with respect to the angle $q$, which is the incident angle of the beam projected onto unit cell plane (Fig. \ref{fig:schematic_unitcell}). In Eq.~({\ref{eq:fs}}), the pressures $p_{s,n}$ and their derivatives (evaluated at $q=0$) are dimensionless, i.e. normalized per intensity and speed of light. Pressures $p_{s,n}$ are implicitly assumed to vary spatially; when this is not the case, the stiffness coefficients simplify with only the beam intensity under the integral (Eq.~\ref{eq:fs_constant}). 

As we show below, the analytical expressions of Eq.~({\ref{eq:fs}})a-d become a powerful tool for predicting stabilizing behavior and for discovering optimal metasurface configurations. 

\subsection{Dynamics of conventional metasurfaces}
 We first consider an interesting question of when can a planar reflective meta-cone exhibit stabilizing behavior in a light field. The structure, shown schematically in Fig. \ref{fig:physical_cone}, is assumed to refract light equivalently to a reflective surface inclined at an angle $\alpha$. A simple analysis of the reflection off of a tilted surface yields the following expressions for the pressure: $p_s=-\sin(2\alpha)$, $p_n=1+\cos(2\alpha)$, and $\partial p_s /\partial q = 1-\cos(2\alpha)$, $\partial p_n / \partial q = -\sin(2\alpha)$, all evaluated at equilibrium ($q=0$) (see Appendix B for details). 

In addition to lateral stabilization, it is important to optimally use the incident beam power. Specifically, we seek to maximize the longitudinal/pushing force relative to the power of the incident beam. The longitudinal force $F_z$ is given by
\begin{equation}
    F_z = \frac{2\pi}{c} \int_0^{D/2} p_n I(s) s \mathrm{d} s 
\end{equation}
Figure \ref{fig:physical_cone}a shows the total longitudinal force as a function of beam width ($w$) and cone angle ($\alpha$), normalized to $F_0=P_0/c$ where $P_0 = 2\pi \int_0^{\infty} I(s) s \mathrm{d}s$ is the total beam power. By inspection, the maximal longitudinal force is $F_z=2F_0$: this corresponds to an object both wide enough to intercept the full beam power and perfectly specularly reflective to change incident photon momentum from $\hbar k_0 \hat{z}$ to $-\hbar k_0 \hat{z}$. The shaded area in the figure corresponds to the case where the necessary condition for stability $c_{1,2,3}<0$ is violated (Eq. \ref{eq:Cs}). For the case of a Gaussian beam of intensity proportional to $e^{-2r^2/w^2}$, we observe no combination of the cone angle and the beam width that leads to stabilizing dynamics.

\begin{figure}[h]
\centering
    \includegraphics[width=80mm]{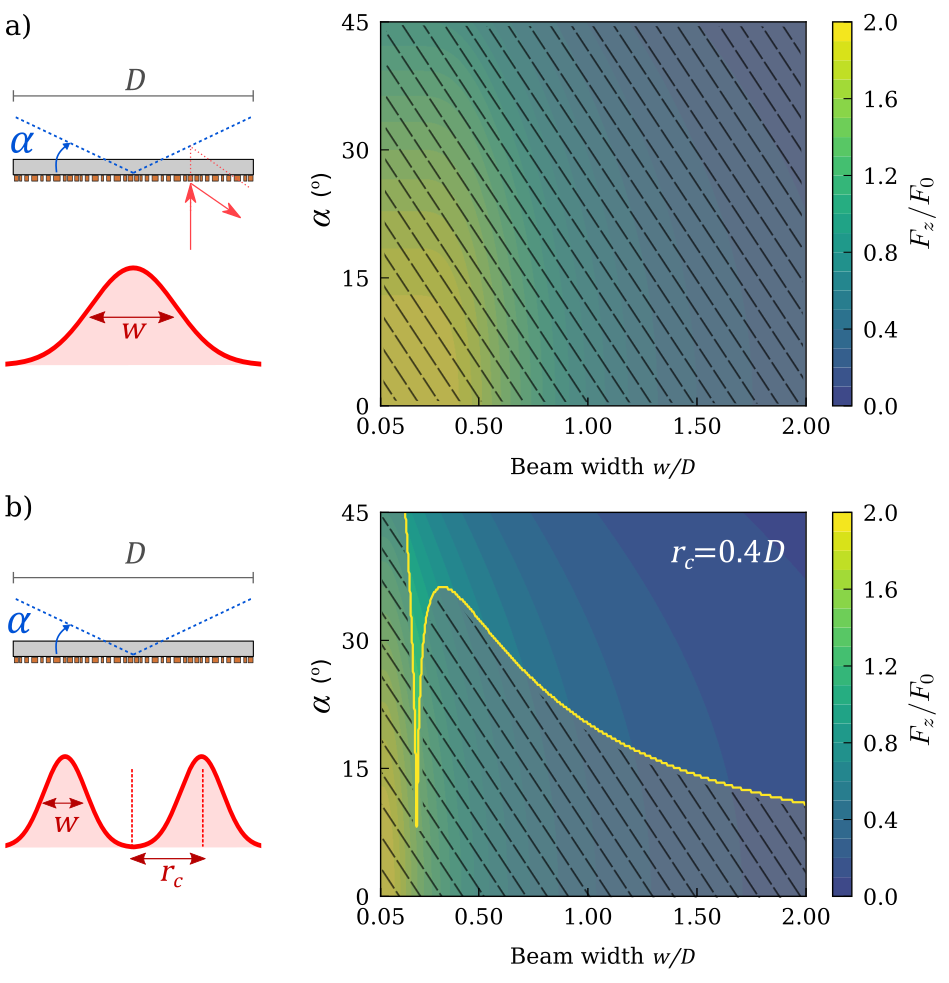}
    \caption{\label{fig:physical_cone} Stability analysis of a planar structure that acts as a reflective meta-cone. (a) Shaded area corresponds to the region of instability, where the necessary condition for stability is violated. For a Gaussian beam intensity profile peaked on axis, no combination of cone angle and beam width can lead to a stable configuration. (b) In an annular beam, candidate stability configurations become possible. All dimensions are normalized to the structure diameter $D$.}
\end{figure}  

An annular incident beam, by contrast, can satisfy the stability conditions (Figure \ref{fig:physical_cone}b). For a beam of the intensity proportional to $e^{-2(r-r_c)^2/w^2}$, we notice a region where stabilization is possible. Interestingly, there appears a trade-off between small cone angle / large beam width (i.e. the photon momentum change is greater but the beam is wider) and large cone angle / small beam width (i.e. momentum change is smaller but the beam is more concentrated on the object). Similarly, we observe a set of candidate solutions near $w \sim 0.2D$. The thin region of candidate configurations that satisfy conditions $c_{1,2,3}$ is associated with a strong variation and sudden change of sign of the rotational stiffness coefficient $\sfc_3$, when a narrow annular beam is concentrated on the outer edges of the structure (analyzed in more detail in the Appendix). For $\alpha<0$, we observe no solutions for either the Gaussian or the annular beam case in Figure \ref{fig:physical_cone}.  

A benefit of choosing the example of a reflective cone is that it allows us to validate our analytical results against a numerical ray tracing model. To verify our analytical expressions, we develop a ray tracing model in which normally incident light scatters off of an inclined reflective surface. The ray-trace model is developed in the the finite-element-method solver COMSOL Multiphysics. Figure~\ref{fig:raytrace}a shows an example of a set of rays, incident from the bottom, scattering off of a tilted cone. For easier visualization, only a handful of rays are shown. In our numerical analysis, we consider $\approx 58,000$ rays to ensure convergence. For each ray, we calculate the momentum change between the incident and the refracted momentum, which enables us to quantify the radiation pressure force.  
To compare against the ray-trace model, we use the previously derived expressions for the radiation pressure components. Because the model in Comsol is a 3D cone (and not a cone-mimicking planar structure), the surface normal and the corresponding projection area factor are modified, namely $ \cos(G) = [k_{s},k_{t},k_{n}] \cdot[-\sin(\alpha),0,\cos(\alpha)] = -k_{s}\sin(\alpha) + k_{n}\cos(\alpha)$. Similarly, because the integration is performed along the cone edge, the limit of radial integration is $D/2 \cos\alpha$. Figure \ref{fig:raytrace}(b,c) shows the comparison between analytical equations and the ray trace model, when the cone angle $\alpha$ and/or the cone rotation angle $\theta$ is varied. The two approaches match to within a fraction of $<0.001$.

\begin{figure}[t]
\centering
    \includegraphics[width=80mm]{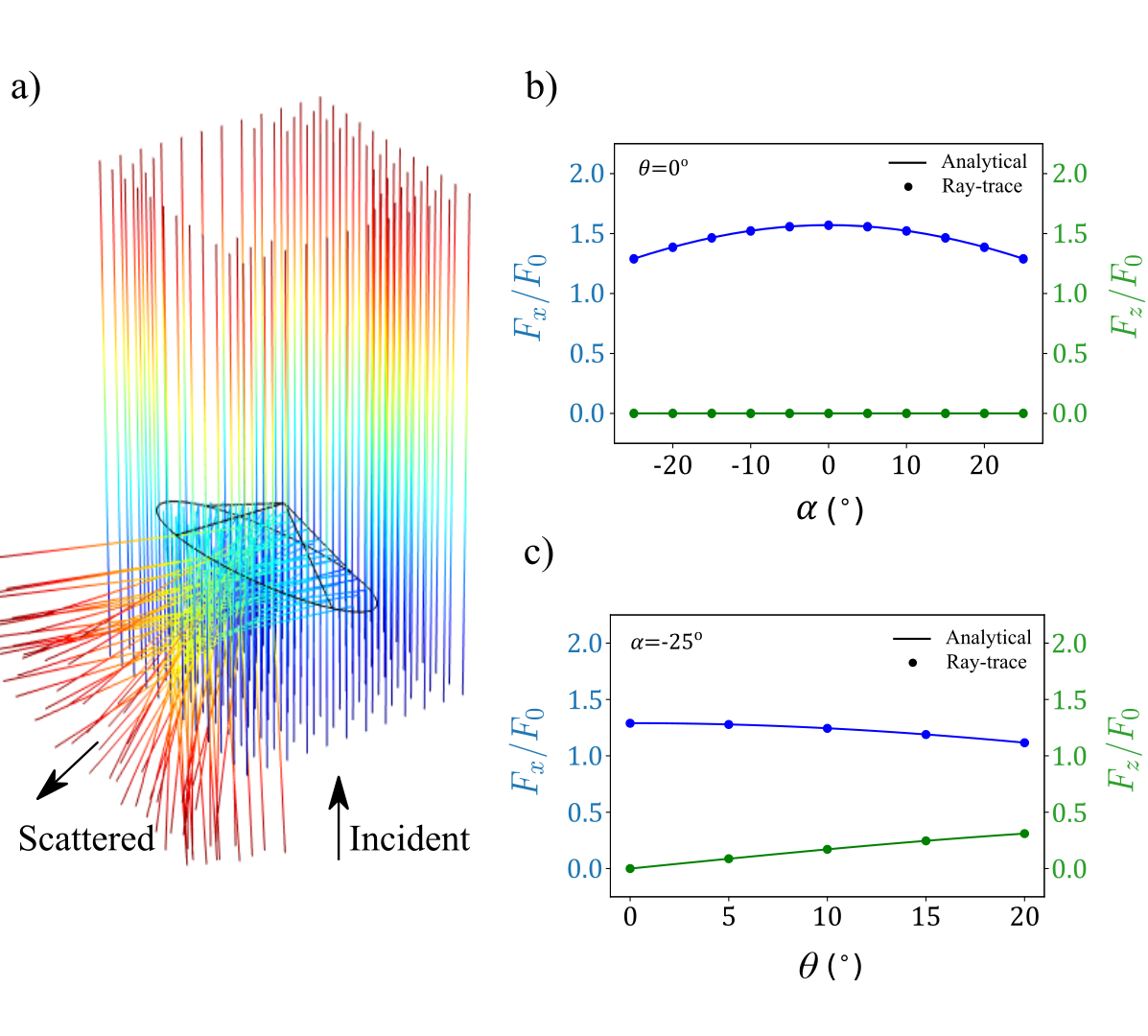}
    \caption{\label{fig:raytrace} Validation of the analytical formalism with a numerical ray-tracing model. (a) Ray trace simulation of light incident on a three-dimensional object. (b) Forces on the cone as the cone angle $\alpha$ is varied. Positive angle represents an inverted cone (apex below the base). The rotation angle $\theta$ and $\phi$ were set to $0^{\circ}$. (c) Forces on the cone as a function of the rotation angle $\theta$, when the cone angle $\alpha$ is constant. The analytical results and the numerical ray-trace simulations in (b) and (c) are indicated with solid lines and circles, respectively.}
    \label{fig:raytrace}
\end{figure}  

We now turn to the discussion of directing phase-gradient metasurfaces, a general family of photonic elements which have attracted significant interest \cite{Genevet_2017}. Typically used for beam-steering, these structures inherently alter the momentum of light in a manner that can be tailored for passively stabilizing optical manipulation. A phase-gradient axial metasurface comprises an array of subwavelength elements that together impart a lateral wave-vector shift $k_s^F = k_s^I + \partial \Phi / \partial s$, where $k^F, k^I$ are the refracted and the incident wave vector, respectively, and s denotes the radial direction.  Figure \ref{fig:metasurface_dynamics}a shows a schematic of such an object where normally incident light is radially directed at an angle $\alpha$, namely $\partial \Phi / \partial s =k_0 \cos(\alpha)$. The expressions for the pressures $p_s(\alpha)$ and $p_n(\alpha)$ are derived in the Appendix A. For generality, we consider $\alpha \in [-\pi, \pi]$ to account for both reflection-mode ($\alpha<0$) and transmission-mode ($\alpha >0$) metasurfaces.

For a metasurface with a constant directing angle, $\alpha(s)=\alpha$, Fig. \ref{fig:metasurface_dynamics}a shows the net longitudinal force versus the beam width and the metasurface angle. The strongest $z$ force is, unsurprisingly, realized for the case of a narrow beam impinging on a back-reflecting structure, a configuration that maximizes the momentum transfer, as indicated in bottom left of Fig. \ref{fig:metasurface_dynamics}a. However, no configuration in that region---or anywhere else for a metasurface in reflection mode---could be stable. The onset of stability candidates is realized when the metasurface directing angle $\alpha$ becomes large, i.e., for metasurfaces that predominantly transmit and not reflect light. Though stabilization could become possible for such transmission-mode metasurfaces, this configuration exhibits weak longitudinal force. This force could be slightly increased for larger metasurface deflection angles, though metasurfaces capable of steering light at large angles with high efficiency can be challenging to realize in practice.
\onecolumngrid
\begin{center}
\begin{figure}[t]
    \includegraphics[width=180mm]{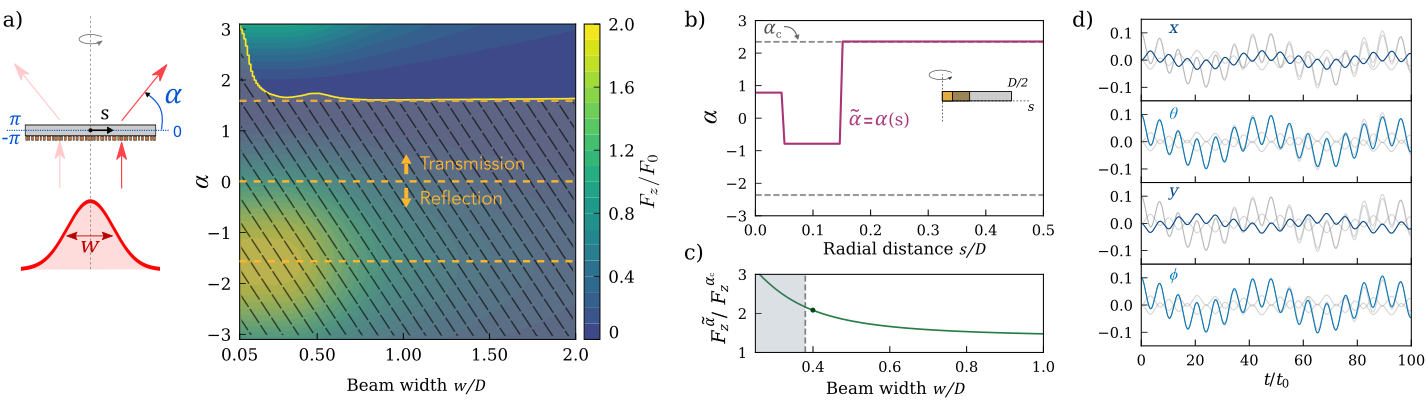}
    \caption{\label{fig:metasurface_dynamics} Stability of a directing metasurface in a Gaussian beam. (a) Shaded region shows instability for constant-angle metasurface configurations except the case of a transparent-mode structure with $\alpha > \pi/2$. In this case, however, the longitudinal $z$ force is weak. (b) A varying radial phase profile $\widetilde{\alpha}\equiv \alpha(s)$ can combine strong optical force of a reflection-mode structure with stabilizing features of a transmission-mode structure. (c) Enhancement of the longitudinal optical force  $F_{z}^{\widetilde{\alpha}}$ relative to the case of constant phase profile $F_{z}^{{\alpha_0}}$ of the same maximum angle $\alpha_{0}$. Shaded area corresponds to the region of instability. (d) Example dynamics for initial perturbation given by $\theta=\phi=0.1$. Here $w/D=0.4$, indicated by a dot in the panel (c). Each panel highlights one coordinate (blue) relative to the three others (shaded gray).}
\end{figure} 
\end{center}
\twocolumngrid
\subsection{Design and optimization of metasurfaces}
A more promising approach to efficient stabilization is to incorporate a radially varying phase profile, so that the strong optical force of a reflecting structure might be combined with the stabilizing features of a transparent structure. In our analysis, we first specify the allowable range of metasurface directing angles $\pm \alpha_c$, shown bounded by the two dashed lines in Fig. {\ref{fig:metasurface_dynamics}}b. For demonstration purposes, here we select $\alpha_c=\frac{3}{4}\pi$, but we note that the choice of $\alpha_c$ could be dictated by external considerations, including fabrication constraints. We then proceed to discretize the phase profile over $N_s$ radial steps (between 0 and $D/2$), where at each step the directing angle can assume one of $N_{\alpha}$ values within the $\pm \alpha_c$ bounds. This results in a total of $(N_s)^{N_{\alpha}}$ combinations. Here, we note there are multiple ways a metasurface object with a target profile can be realized in practice~\cite{Byrnes_2016, Guo2017-wo, Sell2017-za, Phan2019-mr}.

Importantly, we show that the analytical nature of our formalism enables efficient screening of all combinations to obtain optimal solutions that maximize force, while satisfying stability conditions. Following this approach, we obtain the phase profile $\tilde{\alpha}$ which is shown in Fig. \ref{fig:metasurface_dynamics}b (see Appendix B for details). Closer to the center of mass of the metasurface $(s=0)$, we observe non-conventional variation of phase, trending towards smaller values of $\alpha$ with stronger longitudinal force. Elsewhere on the metasurface, we observe that the scattering response appears selected to follow the phase profile at the edge of allowable $\alpha_c$. In Figure \ref{fig:metasurface_dynamics}c, we quantify the enhancement of the force $F_{z}^{\tilde{\alpha}}$ relative to the maximum force for a constant directing angle. Notably, we see that greater than double force $(F_z^{\tilde{\alpha}} / F_z^{\alpha_c} \approx 2.09)$ can be realized for beam widths $w \approx 0.39D$, right at the edge of the region of guaranteed instability (shaded area in Fig. \ref{fig:metasurface_dynamics}c). The enhancement in the magnitude of the $z$ force remains high even for broader beams. The design of optomechanical metasurfaces for a figure of merit other than the longitudinal force, or over a finer phase profile grid is a straightforward extension of the presented analysis.

Once the necessary conditions for stabilizing behavior are satisfied, the dynamics of manipulation can be numerically evolved from the full non-linear equations of motion (Eq.~\ref{eq:EOM}).  For the $\tilde{\alpha}$ phase profile, Figure \ref{fig:metasurface_dynamics}d shows example dynamics of the composite metasurface. The abscissa corresponds to time, normalized in units of $t_0=\sqrt{mc/I_0 D}$ where $I_0$ is the peak beam intensity on the beam axis. The observed dynamics shows strongly coupled translation and rotation: the structure moves along all coordinates, but in a restoring manner. Going beyond this set of initial conditions, as an example of probing a neighborhood around the equilibrium, we sampled all combinations of displacement $x,y=\pm 0.01$ and tilt $\phi,\theta = \pm 0.1$ and observe bounded dynamics over the analyzed timescale ($10^3 t_0$). The dynamics shown in Fig. \ref{fig:metasurface_dynamics}c primarily focuses on lateral stabilization - stabilization relative to the beam axis. However, we note that our treatment also allows for the variation of the beam intensity along the $z$-axis. In such cases, the dynamical behavior of the metasurface would be generally influenced by the beam's Rayleigh range. 

Additionally, the profile of the beam represents a (multi-dimensional) degree of freedom that can be harnessed for effective optomechanical stabilization, and which our framework can easily incorporate. Going beyond a Gaussian beam incident on a metasurface, we analyzed a parametrized envelope applied to the Gaussian intensity profile. We emphasize that the presented formalism lends itself to straightforward calculation of derivatives of the merit function and stability constraints, which enables us to employ efficient gradient-based optimization to show the potential for a substantial additional enhancement ($\approx 170\%$) relative to the case of Fig.~\ref{fig:metasurface_dynamics}b. For the case analyzed in this work, the end result is a design exhibiting a longitudinal optical force that is $12.3\mathrm{x}$ superior to previous examples in the literature \cite{Ilic_2019}. The details of the beam profile analysis are presented in Appendix B.

\section{Discussion and Conclusion}
In conclusion, we have presented an analytical framework for stabilizing manipulation of freestanding photonic metasurfaces in three dimensions. We derived analytical expressions for stiffness coefficients that arise from scattering off of metasurface elements designed to induce restoring forces and torques.  Our investigation shows how the complex three-dimensional optomechanical response is captured by a two-dimensional treatment of the unit cell scattering. The implications of this reduced problem dimensionality are twofold. First, the framework is universally applicable to arbitrary embedded element profiles and/or radial beam variations. Second, our formalism enables efficient design of optomechanical metasurfaces as well as light beam configurations---we show examples of non-conventional phase-gradient profiles and beam intensity variations with substantial (e.g., order-of-magnitude) improvement in performance. These results facilitate the discovery of macroscale photonic objects for stable manipulation in collimated, but otherwise unfocused light beams. Such use of metasurfaces for mechanical stabilization could open up new perspectives for manipulation complementary to traditional optical tweezing, including long-range manipulation and manipulation of macroscopic objects, with potential for terrestrial and space applications \cite{Forward_1984, Johnson_2011, Starshot_2017, Lubin_2016, Atwater_2018, Kulkarni_2018, Parkin_2018, Ilic2020-uq}.

\begin{acknowledgments}
We thank O. Miller for helpful feedback and acknowledge discussions with colleagues from the Breakthrough Starshot Lightsail Initiative. We acknowledge the support from the Minnesota Robotics Institute (MnRI) and acknowledge the Minnesota Supercomputing Institute (MSI) at the University of Minnesota for providing resources that contributed to the research results reported in this paper.\\
\\
\noindent *Email: ilic@umn.edu
\end{acknowledgments}

\appendix
\section{Derivation of the Analytical Formalism for 3D Metasurface Stabilization}

We consider three reference frames of interest: the laser/lab frame ($I$), the frame of the object / body frame ($B$), and the frame of the embedded unit element ($S$)--rotated about $z''$ axis relative to frame $B$. Transformation between the laser frame and the body frame is performed using the 1-2-3 (also known
as $x$-$y'$-$z''$) Euler angle convention of rotations. The direction cosine matrix that transforms a vector from $I$-frame to $S$-frame is given by
\begin{equation}
\begin{aligned}
H_I^S(\beta, \theta, \phi) &=
\left[\begin{matrix}
\cos\beta \cos\theta & 
\cos\beta \sin\theta \sin\phi + \sin\beta \cos\phi \\
-\sin\beta \cos\theta &
-\sin\beta \sin\theta \sin\phi + \cos\beta \cos\phi \\
\sin\theta & -\cos\theta \sin\phi \\
\end{matrix}\right.\\
&\qquad\qquad\left.\begin{matrix}
 -\cos\beta \sin\theta \cos\phi + \sin\beta \sin\phi \\
 \sin\beta \sin\theta \cos\phi + \cos\beta \sin\phi  \\
 \cos\theta \cos\phi
\end{matrix}\right]
\end{aligned}
\end{equation}
The inverse transformation matrix is, by definition, equal to the transpose 
\begin{equation}
    H_S^I = \left(H_I^S\right)^T
\end{equation}
In the lab frame, the wave-vector of incident light is given by $\vec{k}_0 = [0,0,1]$. Transforming to the $S$ frame, it follows that
\begin{gather}
\vec{k}_i^S = H_I^S \vec{k}_0 =   
\begin{bmatrix}
k_s\\
k_t \\
k_n  \\
\end{bmatrix} 
=
\begin{bmatrix}
-\cos\beta \sin\theta \cos\phi + \sin\beta \sin\phi\\
\sin\beta \sin\theta \cos\phi + \cos\beta \sin\phi \\
\cos\theta \cos\phi  \\
\end{bmatrix} 
\label{eq:kS}
\end{gather}
where the incident wavevector components $k_s, k_t, k_n$ (in the $S$ frame) are defined above. The laser beam interacts with the unit cell in the $s$-$n$ plane. The projected incident angle $q$, from Fig. 1 of the manuscript, is then defined as
\begin{equation}
    \tan q = \frac{k_s}{k_n}
    \label{eq:q}
\end{equation}
We also note that the wavevector magnitude in the $s$-$n$ plane is equal to $k_{sn}=\sqrt{k_s^2 + k_n^2} \leq |k_0|$. For a rotated object, we need to take into account the projected area, captured by the cosine of the angle between the surface normal and the incident wavevector $\cos(G)$. Note that, to first order, $\partial\cos(G))/ \partial \theta = \partial\cos(G)/ \partial \phi = 0$, and $\cos(G) = 1$ in equilibrium. 

For analyzing the dynamical nature of motion in the close vicinity of the beam axis, small angles are considered. Under this assumption, the direction cosine transformation matrix becomes
\begin{gather}
 H_{I}^{S} \left (\beta,\theta,\phi \right) = \begin{bmatrix}
\cos (\beta) & \sin (\beta) & -\cos( \beta) \theta + \sin( \beta) \phi\\
-\sin (\beta) & \cos (\beta) & \sin( \beta) \theta + \cos( \beta) \phi \\
\theta & -\phi & 1  \\
\end{bmatrix}
\end{gather}
For small displacements $x,y \ll D$ away from the beam axis, Eq.~(\ref{eq:rbar}) can be approximated
\begin{gather}
\bar{r} =  
\begin{bmatrix}
x + s \cos\beta\\
y + s \sin \beta\\
s(-\theta \cos\beta + \phi\sin\beta)  \\
\end{bmatrix} 
\label{eq:rbar0}
\end{gather}
where the distance to the axis $r_{\perp}^2 \approx (x+s\cos\beta)^2 + (y + s\sin\beta)^2$. Retaining terms up to and including the first order, we further write
\begin{equation}
    r_{\perp} \approx s\sqrt{1 + \frac{2\cos\beta}{s}x + \frac{2\sin\beta}{s} y } \approx s + \cos\beta x + \sin\beta y
\label{eq:rbarD}
\end{equation}
For small rotations, it is similarly true that $q\ll 1$, so from Eq.(\ref{eq:q}), we have:
\begin{equation}
    q = -\theta \cos\beta + \phi \sin\beta
\label{eq:qD}
\end{equation}
We now proceed to evaluate the force and torque terms associated with small perturbations around the equilibrium position on the beam axis. For example,
\begin{gather}
\frac{\partial}{\partial x}
\begin{bmatrix}
F_{x}  \\
F_{y}  \\
F_{z}  \\
\end{bmatrix}^{I}
= \int_{0}^{\frac{D}{2}} ds \int_{0}^{2 \pi} d \beta \ 
\begin{bmatrix}
p_s \cos\beta \\
p_s \sin\beta  \\
p_n   \\
\end{bmatrix}
\frac{\partial I(r_{\perp})}{\partial x} s
\end{gather}
where we recognize that 
\begin{equation}
    \frac{\partial I(r_{\perp})}{\partial x} = \frac{\partial I(r_{\perp})}{\partial r_{\perp}}\frac{\partial r_{\perp}}{\partial x} = \frac{\partial I(s)}{\partial s} \cos \beta = I' \cos\beta
\end{equation}
using the shorthand $I'(s) = \partial I(s) / \partial s$. Similarly, we have
\begin{equation}
    \frac{\partial I(r_{\perp})}{\partial y} = \frac{\partial I(r_{\perp})}{\partial r_{\perp}}\frac{\partial r_{\perp}}{\partial y} = \frac{\partial I(s)}{\partial s} \sin \beta = I' \sin\beta
\end{equation}
Substituting back, we finally obtain
\begin{align}
\frac{\partial F_{x}}{\partial x}  &= \pi \int_{0}^{D/2} ds \ p_{s} \ I'(s) \ s \nonumber \\
 \frac{\partial F_{y}}{\partial x} &= 0 \nonumber \\
 \frac{\partial F_{z}}{\partial x} &= 0
\end{align}
where we utilized the following identities $\int_0^{2\pi} d\beta \cos^2 \beta = \pi$, and $\int_0^{2\pi} d\beta \cos\beta \sin\beta = 0$, $\int_0^{2\pi} d\beta \cos\beta = 0$. 

In a similar fashion, for displacement along the $y$ coordinate, we obtain
\begin{align}
 \frac{\partial F_{x}}{\partial y} &= 0 \nonumber \\
 \frac{\partial F_{y}}{\partial y} &= \pi \int_{0}^{D/2} ds \ p_{s} \ I'(s) \ s \nonumber \\
 \frac{\partial F_{z}}{\partial y} &= 0
\end{align}

An equivalent analysis for the torques gives the following relationships
\begin{gather}
\frac{\partial}{\partial x}
\begin{bmatrix}
\tau_{x}  \\
\tau_{y}  \\
\tau_{z}  \\
\end{bmatrix}
= \int_{0}^{\frac{D}{2}} ds \int_{0}^{2 \pi} d \beta \ 
\begin{bmatrix}
s \sin\beta \ p_{n}  \\
-s \cos\beta \ p_{n}  \\
0   \\
\end{bmatrix}
\frac{\partial}{\partial x}I(r_{\perp})  s
\end{gather}
yielding
\begin{align}
     \frac{\partial \tau_{y}}{\partial x} &= -\pi \int_{0}^{D/2} ds \ p_{n} \ I'(s)\ s^2 \nonumber\\
     \frac{\partial \tau_{x}}{\partial y} &= \pi \int_{0}^{D/2} ds \  p_{n} \ I'(s)\ s^2 \nonumber\\
     \frac{\partial \tau_{x}}{\partial x} &= 0 \nonumber \\ 
     \frac{\partial \tau_{y}}{\partial y} &= 0
\end{align}

Examining small angular displacements $\theta, \phi$ in a similar fashion, it is obtained
\begin{align}
\frac{\partial}{\partial \theta}
\begin{bmatrix}
F_{x}  \\
F_{y}  \\
F_{z}  \\
\end{bmatrix}
= \int_{0}^{\frac{D}{2}} ds &\int_{0}^{2 \pi} d \beta \ 
\frac{\partial}{\partial \theta} I(s) s \\
&\cdot\begin{bmatrix}
p_s \cos\beta + p_n \theta \\
p_s \sin\beta - p_n \phi  \\
p_s(-\theta\cos\beta + \phi\sin\beta) + p_n   \\
\end{bmatrix}
\end{align}
further, to first order this simplifies to
\begin{gather}
\frac{\partial}{\partial \theta}
\begin{bmatrix}
F_{x}  \\
F_{y}  \\
F_{z}  \\
\end{bmatrix}
= \int_{0}^{\frac{D}{2}} ds \int_{0}^{2 \pi} d \beta \ 
\begin{bmatrix}
\frac{\partial{p_s}}{\partial \theta} \cos\beta + p_n \\
\frac{\partial{p_s}}{\partial \theta} \sin\beta  \\
-p_s\cos\beta + \frac{\partial p_n}{\partial \theta}  \\
\end{bmatrix}
I(s) s
\end{gather}

We note that from Eq.~(\ref{eq:qD})
\begin{equation}
    \frac{\partial p_s}{\partial \theta} = \frac{\partial p_s}{\partial q} \frac{\partial q}{\partial \theta} = \frac{\partial p_s}{\partial q} (- \cos\beta)
\end{equation}
leading to
\begin{gather}
\frac{\partial}{\partial \theta}
\begin{bmatrix}
F_{x}  \\
F_{y}  \\
F_{z}  \\
\end{bmatrix}
= \int_{0}^{\frac{D}{2}} ds \int_{0}^{2 \pi} d \beta \ 
\begin{bmatrix}
-\frac{\partial{p_s}}{\partial q} \cos^2\beta + p_n \\
-\frac{\partial{p_s}}{\partial q} \cos\beta\sin\beta  \\
-p_s\cos\beta  -\frac{\partial p_n}{\partial q}\cos\beta  \\
\end{bmatrix}
I(s) s
\end{gather}
After integrating over $d\beta$, the only non-zero term that remains is:
\begin{equation}
    \frac{\partial F_x}{\partial \theta} = \pi \int_{0}^{\frac{D}{2}} ds \left( -\frac{\partial p_s}{\partial q} + 2p_n \right) I(s) s
\end{equation}

Other force/torque gradients are derived in the equivalent manner. We summarize below:
\begin{align}
\label{eq:Fpartials}
\frac{\partial F_{x}}{\partial x} &=  \pi \int_{0}^{D/2} ds \ p_{s}(s)I'(s) s   \nonumber \\
\frac{\partial \tau_{x}}{\partial y} &=  \pi \int_{0}^{D/2} ds \ p_{n}(s)I'(s) s^2   \nonumber \\
\frac{\partial F_{x}}{\partial \theta} &= \pi \int_{0}^{D/2} ds \left[ -\frac{\partial p_{s}(s)}{\partial q} + 2 p_{n}(s) \right]I(s)s \nonumber \\
\frac{\partial F_{y}}{\partial \phi} &= \pi \int_{0}^{D/2} ds \left [ \frac{\partial p_{s}(s)}{\partial q} - 2 p_{n}(s) \right]I(s)s \nonumber \\
\frac{\partial \tau_{x}}{\partial \phi} &= \pi \int_{0}^{D/2}  ds \  \frac{\partial p_{n}(s)}{\partial q}  I(s)s^2   \nonumber \\
\frac{\partial \tau_{y}}{\partial \theta} &= \pi \int_{0}^{D/2}  ds \  \frac{\partial p_{n}(s)}{\partial q}  I(s)s^2  \nonumber \\
\frac{\partial \tau_{x}}{\partial \phi} &= \frac{\partial \tau_{y}}{\partial \theta}  \nonumber \\
\frac{\partial F_{x}}{\partial \theta} &= -\frac{\partial F_{y}}{\partial \phi} \\
\frac{\partial F_{y}}{\partial y} &= \frac{\partial F_{x}}{\partial x} \nonumber \\
\frac{\partial \tau_{y}}{\partial x} &= -\frac{\partial \tau_{x}}{\partial y} \nonumber
\end{align}
where the normalized pressure components (generally spatially dependent) $p_s(s)$ and $p_n(s)$ are evaluated at equilibrium ($q=0$).

\subsection{Dynamical Behavior of a Metasurface}
\noindent For the motion of a metasurface approximated as a rigid body in three dimensions, the equations for the kinematics and the dynamics can be expressed as
\begin{align}
  \label{eq:EOM}
  \dot{\vec{r}} &= \vec{v} \nonumber \\
  \dot{\vec{\alpha}} &= L_B^I \vec{\omega} \nonumber \\
  \dot{\vec{v}} &= \frac{1}{m} \vec{F}(\vec{r}, \vec{\alpha}) \\
  \dot{\vec{\omega}} &=  \mathrm{I}^{-1} \left[ - \vec{\omega}\times
  \mathrm{I}\vec{\omega} + \vec{\tau}(\vec{r}, \vec{\alpha}) \right]
  \nonumber
\end{align}
where $\vec{r}$ is the position and $\vec{\alpha}$ the orientation of the body, and $L_B^I$ is the matrix relating the time derivative of orientation angles to components of the angular velocity. The optical force $\vec{F}(\vec{r}, \vec{\alpha})$ and optical torque $\vec{\tau}(\vec{r}, \vec{\alpha})$ depend on the position and the orientation of the body. Because of axial symmetry $\mathrm{I}_{x}=\mathrm{I}_{y}$, and $\tau_z = 0$. For the dynamics of the system, we characterize the state vector as a set of spatial coordinates ($x, \theta, y, \phi$) and their time derivatives ($\dot{x}, \dot{\theta}, \dot{y}, \dot{\phi}$). The stability analysis describes the response to small perturbations near the origin (i.e. the beam axis). Near the origin, we can linearize Eq.~(\ref{eq:EOM}) by observing $\dot{\theta}=\omega_y$, $\dot{\phi}=\omega_x$, and taking the partial derivatives of $\vec{F}, \vec{\tau}$ with respect to translational and angular displacements. We arrive at the matrix form

\begin{gather}
\frac{d}{dt}
\begin{bmatrix}
 x \\
 \theta \\
 y \\
 \phi \\
  \dot{x} \\
 \dot{\theta} \\
 \dot{y} \\
 \dot{\phi}
\end{bmatrix} = 
\underbrace{
\begin{bmatrix}
0 & 0 & 0 & 0 & 1 & 0 & 0 & 0  \\
0 & 0 & 0 & 0 & 0 & 1 & 0 & 0  \\
0 & 0 & 0 & 0 & 0 & 0 & 1 & 0  \\
0 & 0 & 0 & 0 & 0 & 0 & 0 & 1  \\
\frac{1}{m}\frac{\partial F_{x}}{\partial x} & \frac{1}{m}\frac{\partial F_{x}}{\partial \theta}  & 0 & 0 & 0 & 0 & 0 & 0  \\
\frac{1}{I_{y}}\frac{\partial \tau_{y}}{\partial x}  & \frac{1}{I_{y}}\frac{\partial \tau_{y}}{\partial \theta}  & 0 & 0 & 0 & 0 & 0 & 0  \\
0 & 0 & \frac{1}{m}\frac{\partial F_{y}}{\partial y}  & \frac{1}{m}\frac{\partial F_{y}}{\partial \phi}  & 0 & 0 & 0 & 0  \\
0 & 0 & \frac{1}{I_{x}}\frac{\partial \tau_{x}}{\partial y}  & \frac{1}{I_{x}}\frac{\partial \tau_{x}}{\partial \phi}  & 0 & 0 & 0 & 0  \\
\end{bmatrix}}_{A}
\begin{bmatrix}
 x \\
 \theta \\
 y \\
 \phi \\
  \dot{x} \\
 \dot{\theta} \\
 \dot{y} \\
 \dot{\phi}
\end{bmatrix}
\label{eq:dynsys}
\end{gather}
The stiffness coefficients are explicitly defined as 
\begin{align}
   \sfc_{1} &= -\frac{1}{m} \frac{\partial F_{x}}{\partial x} = -\frac{1}{m} \frac{\partial F_{y}}{\partial y} \nonumber \\
   \sfc_{2} &= \frac{1}{m} \frac{\partial F_{x}}{\partial \theta} = -\frac{1}{m} \frac{\partial F_{y}}{\partial \phi}  \nonumber \\
  \sfc_{3} &= \frac{1}{I_{y}} \frac{\partial \tau_{y}}{\partial x} = -\frac{1}{I_{x}} \frac{\partial \tau_{x}}{\partial y}   \nonumber \\
  \sfc_{4} &= \frac{1}{I_{x}} \frac{\partial \tau_{x}}{\partial \phi} = \frac{1}{I_y} \frac{\partial \tau_{y}}{\partial \theta} \nonumber
\end{align}
Using the previously derived expressions for perturbations in force and torque for small displacements (Eq. \ref{eq:Fpartials}), we arrive at  
\begin{align}
  \sfc_{1} &= -\frac{1}{m} \frac{\partial F_{x}}{\partial x} = -\frac{1}{m} \frac{\partial F_{y}}{\partial y} = -\frac{\pi}{mc} \int_{0}^{D/2} ds \  p_{s} I'(s) s \nonumber \\ 
  \sfc_{2} &= \frac{1}{m} \frac{\partial F_{x}}{\partial \theta} = -\frac{1}{m} \frac{\partial F_{y}}{\partial \phi} = \frac{\pi}{mc} \int_{0}^{D/2} ds \ \left[- \frac{\partial p_{s}}{\partial q} + 2 p_{n}\right ] I(s) s  \label{eq:fs_wc}\\ 
  \sfc_{3} &= \frac{1}{\mathrm{I}} \frac{\partial \tau_{y}}{\partial x} = -\frac{1}{\mathrm{I}} \frac{\partial \tau_{x}}{\partial y} = -\frac{\pi}{\mathrm{I}c} \int_{0}^{D/2} ds \ p_{n}  I'(s) s^2 \nonumber \\ 
  \sfc_{4} &= \frac{1}{\mathrm{I}} \frac{\partial \tau_{x}}{\partial \phi} = \frac{1}{\mathrm{I}} \frac{\partial \tau_{y}}{\partial \theta} = \frac{\pi}{\mathrm{I}c} \int_{0}^{D/2} ds \  \frac{\partial p_{n}}{\partial q}  I(s) s^2  \nonumber
\end{align}
where $I_x=I_y=\mathrm{I}$ and $I(s)$ is the radial intensity of unpolarized beam of light, and the factor of speed of light $c$ is explicitly included. The (degenerate) eigenvalues of the Jacobian matrix are:
\begin{equation}
    \lambda_{1-8} = \pm \frac{1}{\sqrt{2}} \sqrt{(\sfc_{4}-\sfc_{1}) \pm \sqrt{(\sfc_{4}-\sfc_{1})^2+4(\sfc_{1}\sfc_{4}+\sfc_{2}\sfc_{3})}  }
\end{equation}
Due to the symmetry of the eigenvalue expression above, positive and negative eigenvalues would appear in pairs. Seeking to avoid a situation with an exponentially growing solution, it is necessary that all eigenvalues be purely imaginary. For this to be true, the following conditions $c_{1,2,3}$ must be satisfied
\begin{align}
\label{eq:Cs}
c_1 &\equiv \sfc_{4} - \sfc_{1} < 0 \nonumber \\
c_2 &\equiv \sfc_{1}\sfc_{4}+\sfc_{2}\sfc_{3} < 0 \\
c_3 &\equiv -(\sfc_{4}-\sfc_{1})^2-4(\sfc_{1}\sfc_{4}+\sfc_{2}\sfc_{3}) < 0 \nonumber
\end{align}
Assessment of stability is carried out through numerical evaluation of conditions above for any metasurface/beam configuration of interest. To avoid numerical issues associated with comparing very small numbers to zero, when evaluating these necessary conditions we introduce a small offset (0.001). The choice of offset can slightly shift the boundary associated with stability/instability. It is convenient to normalize all lengths and times in the problem in the following manner. Assuming the diameter of the metasurface structure to be $D$, the normalized spatial and temporal coordinates become
\begin{align}
   x &\rightarrow x/{D} \nonumber \\
   t &\rightarrow t/\sqrt{\frac{mc}{I_{0}D}}
\end{align}
where $I_0$ is the (peak) beam intensity. With these in mind, the stiffness coefficients become
\begin{align}
  \sfc_{1} &=   -\pi \int_{0}^{D/2} ds \  p_{s} I'(s) s \nonumber \\ 
  \sfc_{2} &=   \pi \int_{0}^{D/2} ds \ \left[- \frac{\partial p_{s}}{\partial q} + 2 p_{n}\right ] I(s) s \nonumber \\ 
  \sfc_{3} &=  -\frac{\pi}{\gamma}  \int_{0}^{D/2} ds \ p_{n}  I'(s) s^2 \nonumber \\ 
  \sfc_{4} &=  \frac{\pi}{\gamma}  \int_{0}^{D/2} ds \  \frac{\partial p_{n}}{\partial q}  I(s) s^2  
\end{align}
where $\gamma = 1/16$ for a uniform disk. For the case when $p_s, p_n$ are spatially independent, the stiffness expressions further simplify to

\begin{align}
  \sfc_{1} &=   -\pi p_s \int_{0}^{D/2} ds \ I'(s) s \nonumber \\ 
  \sfc_{2} &=   \pi \left[- \frac{\partial p_{s}}{\partial q} + 2 p_{n}\right ] \int_{0}^{D/2} ds \ I(s) s   \label{eq:fs_constant}  \\ 
  \sfc_{3} &=  -\frac{\pi}{\gamma} p_n  \int_{0}^{D/2} ds \ I'(s) s^2 \nonumber \\ 
  \sfc_{4} &=  \frac{\pi}{\gamma} \frac{\partial p_{n}}{\partial q}  \int_{0}^{D/2} ds \ I(s) s^2  \nonumber
\end{align}

\subsection{Metasurface radiation pressure components}
To derive the expressions for the radial $p_s$ and normal $p_n$ pressure components for analyzed metasurface configurations, we begin by considering the net momentum change of the incident beam of light. The total light momentum $\vec{\mathrm{P}}$ can be expressed as $\vec{\mathrm{P}} = \Delta V \vec{g} = \Delta V \ \vec{S}/c^2 = A c \Delta t \ \vec{S} /c^2$, where the momentum density $\vec{g}$ and the $\vec{S}$ Poynting vector relate $\vec{g} = \vec{S}/c^2$, and $A$ is the cross-sectional area.  The force corresponds to the change of momentum $\vec{F} = -\Delta \vec{\mathrm{P}}/ \Delta t$. Assuming the initial wavevector $\vec{k}^I$ and the final wavevector $\vec{k}^F$, the force becomes $\vec{F} = -(\vec{k}^F - \vec{k}^I)/k_0 \ A I_0/c$, where $k_0$ is the wavevector magnitude and $I_0$ is the intensity $I_0=\langle S \rangle$ corresponds to the time-averaged Poynting vector. From here, the unit cell pressure, normalized to $I_0 c$, relates to normalized final/initial wavevectors as $\vec{p} = -(\vec{k}^F - \vec{k}^I)$. In this analysis, it is assumed that the beam intensity is varying slowly relative to the dimension of the unit element.

For the case of a reflective cone, the interaction between the radial unit element and the beam occurs in the $s$-$n$ plane. In this plane, the normalized incident wavevector of the light beam is
\begin{equation}
    \vec{k}^I = \sin(q) \ \hat{s} + \cos(q) \ \hat{n}
\end{equation}
where $q$ is defined by Eq.~(\ref{eq:q}). The wavevector of the specularly reflected beam from Snell's law becomes:
\begin{equation}
    \vec{k}^F = \sin(2\alpha+q) \ \hat{s} - \cos(2\alpha+q)\ \hat{n} \nonumber
\end{equation}
where $\alpha$ is the cone angle. From here, we get the pressure components to be
\begin{align}
 p_{s} &= [-\sin(2\alpha+q) + \sin(q)] \ k_{sn}\nonumber \\
 p_{n} &= [\cos(2\alpha+q) + \cos(q)] \ k_{sn}
\end{align}
At equilibrium ($q=0$), we have $k_{sn}=1$. These expressions and their derivatives become
\begin{align}
 p_{s} &= -\sin(2\alpha) \nonumber \\
 p_{n} &= 1+ \cos(2\alpha) \nonumber \\
 \frac{\partial p_s}{\partial q} &= 1 -\cos(2\alpha)  \\
 \frac{\partial p_n}{\partial q} &= -\sin(2\alpha)  \nonumber
\end{align}

For a metasurface, the normalized incident wavevector of the light beam is similarly:
\begin{equation}
    \vec{k}^I =  \sin(q) \ \hat{s} + \cos(q) \ \hat{n}
\end{equation}
A phase-gradient axial metasurface imparts a radial wave-vector shift $k_s^F = k_s^I + \partial \Phi / \partial s$, namely $k_s^F = (\sin q + \cos \alpha)$. As mentioned in the main text, we consider the range of $\alpha$ to be $\alpha\in [-\pi, \pi]$ so as to capture both reflection-mode metasurfaces ($\alpha<0$) and transmission-mode metasurfaces ($\alpha>0$). Since $|k^F|=|k^I|$, we can write the $n$-component momentum change as $ \Delta k_n = \mathrm{sgn}(\alpha) [ \sqrt{ 1- ( \sin q + \cos \alpha )^2} - \cos q ]$. We can then obtain the pressure components as
\begin{align}
    p_s &= - \cos \alpha \ k_{sn} \nonumber \\
    p_n &= - \left[ \mathrm{sgn}(\alpha)\sqrt{ 1- \left( \sin q + \cos \alpha \right)^2} - \cos q \right ] \ k_{sn} 
\end{align}

At equilibrium ($q=0$), we have $k_{sn}=1$. These expressions and their derivatives become
\begin{align}
 p_{s} &= -\cos(\alpha) \nonumber \\
 p_{n} &= 1- \sin(\alpha) \nonumber \\
 \frac{\partial p_s}{\partial q} &= 0  \\
 \frac{\partial p_n}{\partial q} &= \frac{1}{\tan \alpha}  \nonumber
\end{align}

\section{Optimization of Metasurface and Beam Profiles}

For the optimization of the metasurface design in a Gaussian beam shown in Fig. 3b in the main text, we discretize the phase profile over $N_s$ radial steps (between 0 and $D/2$), where at each step, the directing angle can assume one of $N_{\alpha}$ values within the $\pm \alpha_c$ bounds. Specifically, we analyzed $6^6=46,656$ combinations, where
\begin{equation}
    \alpha \in [-\frac{3}{4}\pi, -\frac{1}{2}\pi, -\frac{1}{4}\pi, +\frac{1}{4}\pi, +\frac{1}{2}\pi, +\frac{3}{4}\pi]
\end{equation}
We note that $\alpha=0$ corresponds to a 90-deg deflection of the beam, which is challenging from a practical point of view, but also not relevant for our purposes (since the longitudinal $z$ force is not particularly strong at $\alpha=0$, as can be seen in Fig. 3a of the main text). 

Following this, we analyze a Gaussian beam intensity profile with a polynomial modification, namely
\begin{align}
    I(r) = (g_0 + g_1r + g_2r^2)^2 e^{-g_3r^2}
\end{align}
where the incident beam intensity is parameterized by a vector $\bar{g} = (g_0,g_1,g_2,g_3)$, and $r$ is the dimensionless radial coordinate normalized to beam diameter. We assume the object scatters according to the $\tilde{\alpha}$ metasurface profile from Fig. 3b of the main text. 

In practice, in order to efficiently use the incident laser power, it is often advantageous to maximize the longitudinal radiation force relative to the total beam power, i.e.
\begin{align}
    \eta = \frac{F_z}{P_0/c} = \frac{\frac{2\pi}{c} \int_0^{D/2} p_n I s \mathrm{d} s}{\frac{2\pi}{c} \int_0^{\infty} I s \mathrm{d}s}
    \label{eq:seff}
\end{align}
subject to constraints corresponding to the necessary conditions for stabilization $c_{1,2,3}<0$ from Eq.~\ref{eq:Cs}. As our formalism lends itself to straightforward calculation of $\partial / \partial \bar{g}$ derivatives of both the merit function $\eta$ and the constraints $c_{1,2,3}$, we employ efficient gradient-based optimization, specifically the method-of-moving-asymptotes (MMA)~\cite{Svanberg_2002} accessed via the NLopt package~\cite{Johnson_2017}. Starting from the initial case of a Gaussian beam of Fig. 3 (for which $\bar{g}=(1,0,0,12.5)$), we find a beam intensity profile specified by $\bar{g}=(0,21.,-62.,40.)$. For practical considerations, we sought to make the gradient of beam intensity to be zero on axis. The corresponding enhancement due to the described beam profile optimization is $\eta_f/\eta_i = 1.71$, leading to the overall value of $\eta=0.995$ for the example configuration considered in this work.

By contrast, calculating the same longitudinal force per unit beam power (given by Eq. \ref{eq:seff}) for the structure of Ref. \cite{Ilic_2019}, yields $\eta=0.081$, a consequence of suboptimal photonic structure/beam configuration (Figs. 3/4 in \cite{Ilic_2019}). Taking the ratio of the two $\eta$ values gives the relative improvement of $\approx$12.3x. We note that the structure of this work can be further improved by extending the self-imposed choice of allowable range of angles beyond $[-\frac{3\pi}{4}, \frac{3\pi}{4}]$. This analysis represents one example of how the beam and structure degrees of freedom can be engineered and optimized for stabilization.

%

\end{document}